\documentclass[12pt]{article}
\usepackage{times}
\usepackage{fullpage}
\usepackage{cite}

\usepackage{amsmath}
\usepackage{graphicx}
\usepackage{url,epsfig,subfigure}

\newcommand{\ones}{\mathbf 1}
\newcommand{\reals}{{\mbox{\bf R}}}

\newcommand{\cf}{{\it cf.}}
\newcommand{\eg}{{\it e.g.}}
\newcommand{\ie}{{\it i.e.}}

\newcommand{\BEAS}{\begin{eqnarray*}}
\newcommand{\EEAS}{\end{eqnarray*}}
\newcommand{\BEA}{\begin{eqnarray}}
\newcommand{\EEA}{\end{eqnarray}}
\newcommand{\BEQ}{\begin{equation}}
\newcommand{\EEQ}{\end{equation}}
\newcommand{\BIT}{\begin{itemize}}
\newcommand{\EIT}{\end{itemize}}

\newcommand{\figurewidth}{0.75\columnwidth}
\graphicspath{{figures/}}
\allowdisplaybreaks 

\title{Network Optimization for Unified Packet and Circuit Switched Networks}
\author{Ping Yin\thanks{Ping Yin is with the Department of Electrical and Computer Engineering at the University of California, San Diego, La Jolla, CA 92093-0407. Email: piyin@eng.ucsd.edu.}
\and
Steven Diamond\thanks{Steven Diamond is with the Department of Computer Science at Stanford University, Stanford, CA 94305. Email: diamond@cs.stanford.edu.}
\and
Bill Lin\thanks{Bill Lin is with the Department of Electrical and Computer Engineering at the University of California, San Diego, La Jolla, CA 92093-0407. Email: billlin@eng.ucsd.edu.}
\and
Stephen Boyd\thanks{Stephen Boyd is with the Department of Electrical Engineering at Stanford University, Stanford, CA 94305. Email: boyd@stanford.edu.}}
\date{}

\begin{document}
\maketitle

\begin{abstract}
Internet traffic continues to grow relentlessly, driven largely by increasingly high resolution video content. Although studies have shown that the majority of packets processed by Internet routers are pass-through traffic, they nonetheless have to be queued and routed at every hop in current networks, which unnecessarily adds substantial delays and processing costs. Such pass-through traffic can be better circuit-switched through the underlying optical transport network by means of pre-established circuits, which is possible in a unified packet and circuit switched network. In this paper, we propose a novel convex optimization framework based on a new destination-based multicommodity flow formulation for the allocation of circuits in such unified networks. In particular, we consider two deployment settings, one based on real-time traffic monitoring, and the other relying upon history-based traffic predictions. In both cases, we formulate global network optimization objectives as concave functions that capture the fair sharing of network capacity among competing traffic flows. The convexity of our problem formulations ensures globally optimal solutions.
\end{abstract}

\section{Introduction}


Internet traffic continues to grow unabatedly at a rapid rate, driven largely by more and more video content, from 1080p HD to 4K Ultra HD video streaming today, to 8K Ultra HD video streaming in the near future. Although the packet-switching approach used in Internet backbone networks has thus far been able to keep up, it is unclear whether electronic routers that have been used at the core of backbone networks will continue to scale to match future traffic growth. On the other hand, optical fiber and switching elements have demonstrated an abundance of capacity that appears to be unmatched by electronic routers. The rate of increase in optical transport capacity has been keeping pace with traffic growth. Thus, one way of keeping pace with future traffic demands is to build an all-optical backbone network. However, packet switching requires the buffering of packets, of which optical switches are not capable today, and it appears unlikely that reasonable size packet buffers can ever be practically realized in optics. On the other hand, circuit switching has a much simpler data transport, making it well-suited to optics and its vast capacity potential.


To harness the huge capacity of optical circuit switching in IP networks, researchers have explored different ways of implementing IP over dynamically configurable optical transport networks \cite{coplar,banerjee01,yoo03,sengupta03,li07,morrow05,GMPLS00}. These earlier efforts
assumed a GMPLS-based control plane \cite{morrow05,GMPLS00}. More recently, given the broad success of software-defined networking (SDN) \cite{shenker11,kreutz15,xia15,mckeown08}, there has been considerable renewed interest in unified packet and circuit switched network architectures based on SDN as the unified control plane \cite{das12,das13}.
In the SDN-based unified architecture proposed in \cite{das13},
backbone routers are replaced with less expensive hybrid optical-circuit/electrical-packet switches that have both circuit-switching and packet-switching capabilities. These hybrid switches are logically connected in a fully-meshed network where each hybrid switch implements an IP node, and where each IP node is logically connected to each and every other IP node via a single direct circuit-switched hop. This unified packet and circuit-switched network can then be managed using a single converged control plane.

\begin{figure}[t]
    \centering
    \includegraphics[width=\columnwidth]{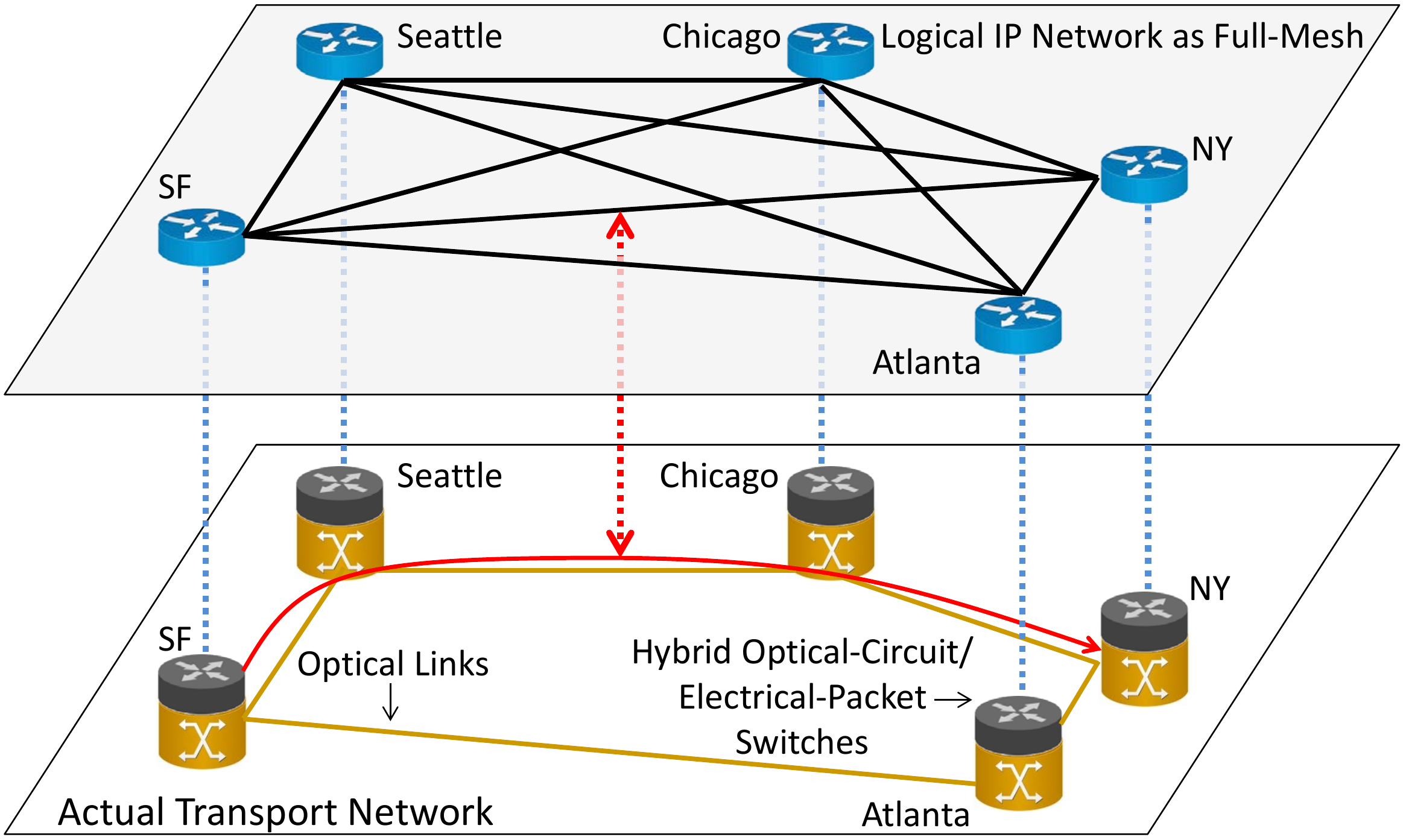}
    \caption{IP network logically as a full-mesh, with logical connections
             implemented over an optical circuit-switched transport network and logical
             routers implemented as part of hybrid optical-circuit/electrical-packet switches.}
    \label{fig:IPmesh}
\end{figure}

Figure~\ref{fig:IPmesh} depicts this unified fully-meshed IP network architecture. The actual underlying optical transport network can be dynamically allocated to provide different circuit capacities to implement each logical connection in the full-mesh, for example based on estimated traffic demands. For example, in figure~\ref{fig:IPmesh}, a logical connection from San Francisco (SF) to New York (NY) may be implemented as an optical circuit-switched path via Seattle and Chicago. In general, a logical connection may be implemented over multiple physical paths.

There are several key advantages with an SDN-based unified fully-meshed architecture:
\begin{itemize}
\item First, studies have shown that up to 85\% of the packets that are processed by backbone routers today are just pass-through traffic \cite{simmons14,sengupta03,das13}. Therefore, packets are unnecessarily delayed due to queuing time at intermediate routers. With a unified architecture, packets can traverse the circuit-switched network through pre-established circuits (light-paths) at optical speeds from the source node to the destination node in a single logical hop.

\item Second, backbone routers are unnecessarily expensive today because they must be designed to process all packets, including all pass-through packets. With a unified architecture, expensive packet-switched router ports are primarily needed only for interfacing with access routers; pass-through traffic can be handled by less expensive circuit-switched ports. This approach promises to dramatically reduce capital expenditures \cite{simmons14,sengupta03,li07,huelsermann08,das13}.

\item Finally, a unified architecture is expected to be far more scalable since most traffic can be switched end-to-end using scalable optical transports.
\end{itemize}

A key problem that must be solved in this unified architecture approach is the allocation of optical circuits between adjacent IP nodes in the logical full-mesh (\ie, between every IE pair of ingress and egress nodes). In this paper, we propose to formulate our circuit allocation problems as convex optimization problems. In particular, the main contributions of this work are as follows:
\begin{itemize}
\item We propose a novel convex optimization framework based on a new destination-based multicommodity flow formulation for the allocation of circuits in unified packet and circuit switched networks.

\item We consider two deployment settings for circuit allocation. In the first setting, we consider the case in which real-time traffic measurements are possible, and we can dynamically allocate circuits on a frequent basis in response to changing traffic.

\item In the second setting, we consider the case in which we allocate circuits based on historical traffic patterns. Previous studies have shown that the aggregate traffic at the core of the network tends to be very smooth and that it follows strong diurnal patterns \cite{coplar,roughan03,medina02}. Such diurnal traffic observations over repeated data sets suggest that circuits can be allocated based on historical data. In this setting, circuit configurations can be precomputed offline.

\item In both settings, we formulate global network optimization objectives as concave functions that capture the fair sharing of network capacity among competing traffic flows.  The convexity of our problem formulations ensures globally optimal solutions.
\end{itemize}

The rest of this paper is organized as follows. In \S\ref{sec:flow}, we present our destination-based multicommodity flow formulation for circuit allocation that reduces the number of decision variables in the convex optimization problems by a factor of $n$, where $n$ is the number of nodes in the network. In \S\ref{sec:allocate}, we present our formulations of the real-time based and history-based circuit allocation problems as convex optimization problems. We then describe our experimental setup in \S\ref{sec:setup}, and we present the results of our evaluations in \S\ref{sec:results}. In \S\ref{sec:related}, we discuss additional related work.
Finally, we present concluding remarks in \S\ref{sec:conc}.

\section{Destination-Based Multi-Commodity Flow Formulation}
\label{sec:flow}

We formulate our optical circuit allocation problems as multi-commodity flow optimization problems.  We consider a backbone network with $n$ nodes and $m$ directed edges, and we index nodes as $i=1,\ldots, n$ and edges as $j=1,\ldots,m$. Note that an undirected edge between nodes $k$ and $\ell$ can be modeled by two directed edges, one from $k$ to $\ell$ and the other from $\ell$ to $k$. With $n$ nodes, we have $n(n-1)$  ingress-egress (IE) pairs, and we index IE pairs as $(k,\ell)$, which refers to ingress (source) $\ell$ and egress (destination) $k$ (\ie, from node $\ell$ to node $k$).

Classically, multi-commodity flow formulations typically use $n(n-1)m$ flow assignment variables, each of which defines the fraction of the corresponding IE pair traffic (among $n(n-1)$ IE pairs) along the corresponding edge (among $m$ edges).  In this paper, we propose a \emph{destination-based} multi-commodity flow formulation in which the \emph{flows} (``commodities'') are labeled by their destination or egress node $k$ rather than by an IE pair.  This reduces the number of flow assignment variables by a factor of $n-1$ to $nm$ variables.  This substantial reduction in the number of variables allows us to scale the method to far larger networks.   To the best of our knowledge, our proposed compact formulation has not been proposed before in networking.  This destination-based multi-commodity flow formulation is described in the remainder of this section.   We then describe our optimization objectives as concave functions in the subsequent sections so that the optimization problems can be solved with convex optimization.

\paragraph{Traffic demand matrix.}

We denote the \emph{traffic demand} from node $\ell$ to node $k$ as $T_{k\ell}\geq 0$.  Accordingly, we refer to the corresponding $n \times n$ matrix $T$ as the \emph{traffic demand matrix}.  As a node $k$ has no traffic to itself that requires transport on the network, we conveniently redefine $T_{kk}$ to be
\[
	T_{kk}=-\sum_{\ell\neq k} T_{k\ell},
\]
the negative of the total traffic demand for, and exiting at, node $k$.  With this definition of $T_{kk}$, we have
\[
	\sum_\ell T_{k\ell}=0,
\]
\ie, $T\ones=0$, where $\ones$ is the vector with all entries equal to one.  As defined, $T$ is a Metzler matrix.  Note that the traffic matrix $T$ gives us the IE pair traffic (the $n(n-1)$ off-diagonal entries $T_{k\ell}$, $k\neq \ell$) as well as the total traffic demand for each of the $n$ nodes ($-T_{kk}$).

\paragraph{Multi-commodity flow conservation.}
The traffic flows from ingress node to egress node over a network with $m$ directed edges, as described by its incidence matrix $A\in \reals^{n \times m}$, where
\[
	A_{ij}= \left\{\begin{array}{rl}
		+1 & \mbox{if edge $j$ enters node $i$}\\
		-1 & \mbox{if edge $j$ leaves node $i$}\\
		 0 & \mbox{otherwise.}
	\end{array}\right.
\]
We assume that the network is completely connected, \ie, there is a directed path from any node to any other, which is typically the case for backbone networks.

We allow the splitting or aggregation of network flows that are destined to the same egress node.  Let $F_{kj}\geq 0$ denote the \emph{flow} on edge $j$ that is destined for destination $k$.  As mentioned, this is a multi-commodity flow problem, with $n$ different flows
labeled by their destination or egress node $k$.

At each node, and for each of the $n$ flows, we must have flow conservation, taking into account the ingress flow at the node, the flow entering the node from incoming edges, the flow leaving the node over outgoing edges, and (when the node is the egress node) the flow egressing from the node.  For a node $i \neq k$ (\ie, not the egress node), the ingress flow plus the net flow into the node must sum to zero:
\BEQ\label{e-flow-cons}
	T_{ki} + \sum_j A_{ij}F_{kj} = 0, \quad i, k=1, \ldots, n, \quad i \neq k.
\EEQ
By net flow, we mean the sum of flows entering on incoming edges minus the sum of the flows leaving on outgoing edges.  At the destination node, all the traffic exits, so we have
\BEQ\label{e-flow-cons-self}
	\sum_j A_{ij}F_{ij} = \sum_\ell T_{i\ell}= -T_{ii}, \quad i=1, \ldots, n.
\EEQ
The lefthand side is the net flow into node $i$, and the righthand side is the total of all traffic exiting the network at node $i$.  Equation~(\ref{e-flow-cons-self}) is identical to~(\ref{e-flow-cons}) for $k=i$.  So~(\ref{e-flow-cons}) holds for all $i,k=1,\ldots, n$.  In fact, the $n$ equalities~(\ref{e-flow-cons-self}) hold automatically, which can be seen by summing (\ref{e-flow-cons}) over all edges, so they are redundant.  Therefore, we can simply express multi-commodity flow conservation in a compact matrix formula as
\BEQ\label{e-multiflow-cons}
	T+FA^T=0.
\EEQ

\paragraph{Edge capacities.}
The total traffic on edge $j$ is
$\sum_k F_{kj}$.
In the simplest model, each edge has a capacity that cannot be exceeded, \ie,
\BEQ \label{e-cap}
	\sum_k F_{kj} \leq c_j, \quad j=1, \ldots, m,
\EEQ
where $c_j$ is the \emph{capacity} of edge $j$.  This can be written as $F^T \ones \leq c$, where the inequality is elementwise.


\paragraph{Feasible traffic demands.}
A traffic demand matrix $T$ (with $T_{k\ell}\geq 0$ for $k\neq \ell$ and $T\ones =0$) can be supported by the network if there exists $F\geq 0$ for which (\ref{e-cap}) and (\ref{e-flow-cons}) hold, \ie,
\BEQ\label{e-feas-traffic}
	F \geq 0, \qquad T+FA^T =0, \qquad F^T\ones \leq c.
\EEQ
This set of inequalities, together with $T_{k\ell} \geq 0$ for $k\neq i$ and $T\ones =0$, defines a polyhedron, which we denote as $\mathcal T$.  We refer to a traffic demand matrix $T$ as feasible if $T\in \mathcal T$ (\ie,  a feasible traffic demand matrix is one for which there is a set of edge flows that respects flow conservation and edge capacities).


\section{Formulation of Circuit Allocation Problems}
\label{sec:allocate}

\subsection{General Approach}

To formulate our circuit allocation problems as convex optimization problems, we define a utility function $\phi_{k\ell}(T_{k\ell})$ for each IE pair $(k, \ell)$, $k \neq \ell$, that computes the \emph{utility} of allocating a circuit with capacity $T_{k\ell}$ to IE pair $(k, \ell)$ (\ie, a circuit that can support traffic demand up to $T_{k\ell}$).  We use the compact notation $\phi(T)$ to denote the matrix with entries $\phi_{k\ell}(T_{k\ell})$ when $k \neq \ell$, and we set the diagonal entries of $\phi(T)$ to one.  As discussed below, for both the real-time-based and history-based circuit allocation formulations, $\phi_{k\ell}(T_{k\ell})$ is defined (and required) to be an increasing concave function.

To fairly allocate network resources to implement circuits for different IE pairs, we use the well-known utility fairness notion called $\alpha$-fairness \cite{alphafairness}, which is defined as follows:
\BEQ\label{e-alpha}
U(f) = \left\{\begin{array}{ll}
	\frac{f^{1-\alpha}}{1-\alpha} & \mbox{for $\alpha \geq 0$ and $\alpha \neq 1$}\\
	\log f & \mbox{for $\alpha = 1$}
\end{array}\right.
\EEQ
Depending on the choice of $\alpha$, different notions of fairness can be achieved.  For example, maximum utility is obtained when
$\alpha = 0$, proportional fairness is obtained when $\alpha \rightarrow 1$, and max-min fairness is obtained when $\alpha \rightarrow \infty$.  In practice, a large $\alpha$ is sufficient to ensure max-min fairness.  For any $\alpha \geq 0$, $U(f)$ is an increasing concave function.  We then formulate the circuit allocation problem as follows:
\[
\begin{array}{ll}
\mbox{maximize} & \sum_{k,\ell} U(\phi_{k\ell}(T_{k\ell}))\\
\mbox{subject to} & T \in \mathcal T,
\end{array}
\]
where $T \in \mathcal T$ corresponds to the set of constraints defined in~(\ref{e-feas-traffic}).  We refer to the objective as the \emph{total network utility}.  Since an increasing concave function of a concave function is still concave \cite{boydbook}, $U(\phi_{k\ell}(T_{k\ell}))$ is a concave function of $T_{k\ell}$.  The objective is a sum of concave functions, and therefore it is also a concave function.  Maximizing a concave function subject to convex constraints (\ie, linear equality and inequality constraints) is a convex optimization problem.

Since the objective is an increasing function of $T$, we see that at the optimal point, all edge traffic will actually be equal to the edge capacity (\ie, we will have $F^T\ones =c$).  Therefore, we can replace the inequality $F^T\ones \leq c$ in~(\ref{e-feas-traffic}) with the equality constraint $F^T\ones =c$.  The convex optimization problem then becomes
\BEQ\label{e-network-opt}
\begin{array}{ll}
\mbox{maximize} & \sum_{k,\ell} U(\phi_{k\ell}(T_{k\ell}))\\
\mbox{subject to} & F\geq 0, \\
& T+FA^T=0, \\
& F^T\ones =c.
\end{array}
\EEQ
with variables $T$ (the traffic demands that can be supported) and $F$ (the detailed network flows).

In the remainder of this section, we consider two versions of the circuit allocation problem.  In the first case, we consider the deployment setting in which \emph{real-time traffic measurements} are possible, and we can dynamically allocate circuits on a frequent basis in response to changing traffic.  In the second case, we consider the deployment setting in which we allocate circuits based on \emph{historical traffic patterns}.  In both versions of the problem, we optimize for \emph{utility max-min fairness} by using a sufficiently large $\alpha$ value in~(\ref{e-alpha}).  The two problems differ in how we define the utility functions $\phi_{k\ell}(T_{k\ell})$ for the IE pairs.

\subsection{Real-Time-Based Allocation}
\label{sec:realtime}

In this section, we consider the deployment setting in which actual traffic can be measured in real-time at a reasonable timescale, and that circuits can be dynamically reconfigured.  In particular, let $r_{k\ell}$ be the traffic rate at the current measurement interval for IE pair $(k, \ell)$.  Intuitively, the traffic pattern for the next time interval should be similar to the current measurement interval if the measurement/reconfiguration interval is sufficiently short.  Therefore, we aim to allocate circuit capacities proportional to the current traffic rates, but we want to fully allocate all network resources even when circuit allocations cannot be further increased for some IE pairs.  In particular, we define
\BEQ\label{e-realtime-util}
	\phi_{k\ell}(T_{k\ell}) = \frac{T_{k\ell}}{r_{k\ell}}
\EEQ
By defining the utility function $\phi_{k\ell}(T_{k\ell})$ this way, the solution to network optimization problem~(\ref{e-network-opt}) corresponds to the \emph{weighted max-min fair} solution.

\subsection{History-Based Allocation}
\label{sec:historical}

Alternatively, in this section, we consider the deployment setting in which real-time traffic measurements are not possible.  In this case, we make use of historical traffic statistics to predict the traffic demands for a given time period.  Previous studies have shown that the aggregate traffic at the core of the network tends to be very smooth and that it follows strong diurnal patterns.  In particular, historical traffic demands during a particular time of day (\eg, 11:00-11:30am on a weekday) are a good indicator of expected future traffic demands over the same time of day.  Let $\mbox{\bf r}_{k\ell} = \{r_{k\ell}(1), r_{k\ell}(2), \ldots, r_{k\ell}(t)\}$ be a collection of $t$ historical traffic measurements taken at a particular time of day for the IE pair $(k, \ell)$.  The corresponding empirical cumulative distribution function (CDF) $\Phi_{k\ell}: \reals_+ \to [0,1]$ maps a circuit capacity $T_{k\ell}$ (\ie, the amount of traffic demand that the circuit can support) to the fraction of $\mbox{\bf r}_{k\ell}$ data points that can be supported:
\BEQ\label{e-cdf}
\begin{aligned}
	\Phi_{k\ell}(T_{k\ell}) &=& \frac{\#measurements \leq T_{k\ell}}{t} \\
	&=& \frac{1}{t} \sum_{i=1}^t I[r_{k\ell}(i) \leq T_{k\ell}]
\end{aligned}
\EEQ
where $I[r_{k\ell}(i) \leq T_{k\ell}]$ is the indicator that the measurement $r_{k\ell}(i)$ is less than or equal to the circuit capacity $T_{k\ell}$.

From an empirical CDF $\Phi_{k\ell}(T_{k\ell})$, we derive an increasing concave function $\phi_{k\ell}(T_{k\ell})$ by curve fitting the empirical CDF.  That is, for each data point $r_{k\ell}(i) \in \mbox{\bf r}_{k\ell}$, we have the corresponding empirical CDF data point $\Phi_{k\ell}(r_{k\ell}(i))$.  In general, $\Phi_{k\ell}(T_{k\ell})$ is not concave.  However, for traffic values above the median measured data rate, the corresponding probability density function (PDF) of traffic is typically decreasing, which is reasonable to assume.  Therefore, we simply curve fit $\phi_{k\ell}(T_{k\ell})$ to all the empirical CDF data points above the median data rate in $\mbox{\bf r}_{k\ell}$ (\ie, for all $\Phi_{k\ell}(r_{k\ell}(i))$ such that $r_{k\ell}(i) \geq \mbox{median}(\mbox{\bf r}_{k\ell})$) using an increasing concave functional form.  In general, any increasing concave function can be used as the parametric form for curve fitting.  As we shall see in \S\ref{sec:model}, we have found that an increasing concave \emph{piecewise linear} (PWL) functional form can accurately approximate the empirical CDFs above the corresponding median historical data rate. As another example, fitting the historical data rates to a log-concave functional form would be another natural way to accurately approximate the empirical CDFs.

By deriving the utility function $\phi_{k\ell}(T_{k\ell})$ from the empirical CDF of the historical traffic, we are maximizing the \emph{probability} that the allocated circuits can handle future traffic demands if future traffic demands follow similar traffic patterns as the measured historical traffic.  Correspondingly, the solution to the network optimization problem~(\ref{e-network-opt}) corresponds to the \emph{utility max-min fair} solution where the utility function is derived from historical traffic.

\subsection{Deriving Per-IE Pair Circuit Configurations}

By solving for the convex optimization problem~(\ref{e-network-opt}) with the utility functions defined in either \S\ref{sec:realtime} or \S\ref{sec:historical}, we know what circuit capacities $T_{k\ell}$ can be realized for each IE pair $(k, \ell)$.  However, in our destination-based multi-commodity flow formulation, a \emph{flow} corresponds to all traffic that are destined for the same destination $k$, and the flow assignment variables $F_{kj} \geq 0$ denote the flow on edge $j$ that is destined for destination $k$.  As discussed earlier, this formulation enables us to reduce the number of variables by a factor of $n - 1$, which allows us to scale our approach to far larger networks.

To derive the actual circuit configurations on a per-IE pair basis, we have to disaggregate a single destination flow into parts associated with the different IE pairs.  This has nothing to do with the optimization method, and does not affect what traffic profiles that we are able to support.

As observed earlier, the solution must satisfy $F^T\ones = c$ (\ie, all the edge capacity is used).  Given this constraint, we can show that for each flow with a given destination, there are no (directed) cycles.
To see this, suppose that for destination $k$ there is a nonzero (\ie, positive) directed cycle.  This means there are edges $e_1, \ldots, e_p$ that form a directed cycle, and the flow destined for node $k$ is positive on each of these edges.  This implies that we can reduce the flow destined for node $k$ on each of these edges by some positive amount, and remain feasible. By reducing the flow on each of these edges, we now have unused capacity on these edges, which we can use by assigning (for example) to the IE pairs associated with those edges.  This increases these IE pair traffic values, which increases the objective, which shows the original flow was not optimal.  Therefore, the optimal solution contains no (directed) cycles for each destination flow.  We can exploit this property in deriving the per-IE pair circuit configurations.

In particular, we start with the traffic matrix $F$, which gives the flow on each edge for each destination.  Our goal is to give a more detailed flow description $Z_{k\ell,j} \geq 0$, which is the flow on edge $j$ for the IE pair $(k,\ell)$.  For each IE pair $(k,\ell)$, the edges with nonzero
$Z_{k\ell,j}$ values show us the route or routes that IE pair $(k,\ell)$ takes.  This must satisfy the obvious flow conservation, where it is conserved for all nodes other than $k$ or $\ell$, the traffic $T_{k\ell}$ enters at node $\ell$ and leaves at node $k$.  These detailed flows must satisfy $Z_{k\ell,j}\geq 0$ and
\[
\sum_\ell Z_{k\ell,j} = F_{kj},
\]
but otherwise are completely arbitrary. We describe two simple methods for constructing $Z$ given $F$, but many other methods could be used as well.

Indeed, any method that attributes flow to each IE pair $(k,\ell)$ such that the remaining
flow satisfies all the conditions described above (though with $T_{k\ell}$ set
to zero) will work. The lack of flow cycles ensures that all flow 
can be attributed to IE pairs.


\paragraph{Greedy assignments.}
Consider an IE pair $(k,\ell)$.  We can route the traffic from node $\ell$ to node $k$ in a greedy way.  Starting at node $\ell$, route all the flow along an outgoing edge $j$ with $F_{kj} \geq T_{k\ell}$, if there is such an edge.  If there is no such edge, we will need to split the flow into two or more edges.  We repeat this until we get to the destination.  Then we subtract these flows from the $F$ matrix, which leaves the flows destined for node $k$, other than the flow originating at node $\ell$.  We repeat the process.  This method always works; it tends to avoid splitting flows.


\paragraph{Proportional assignments.}
Alternatively, we can route the traffic for IE pair $(k,\ell)$ from node $\ell$ by splitting the flow proportionally across outgoing edges $j_h$ to a node $h$ that have $F_{kj_h} > 0$. Our multi-commodity flow formulation ensures that
\[
	\sum_h F_{kj_h} \geq T_{k\ell}.
\]
In particular, we assign to the detailed flow
\BEQ\label{e-proportional-split}
	Z_{k\ell,j_h} = T_{k\ell} \left( \frac{F_{kj_h}}{\sum_h F_{kj_h}} \right), \quad \mbox{for each } j_h.
\EEQ
We repeat this by proportionally splitting each $Z_{k\ell,j_h}$ across the outgoing edges of node $h$ until we get to the destination.  Like the greedy assignment method, we subtract these detailed flows from the $F$ matrix.  We repeat this process for other IE pairs.  This method also always works. It tends to split the flows a lot; more specifically, whenever a flow splits at a node, then all IE pairs will also split there.

\section{Evaluation Setup}
\label{sec:setup}

\subsection{Network and Traffic Matrices}
\label{sec:traffic}

We have evaluated our proposed network optimization framework on the optimal circuit allocation problem on a real, large PoP (point of presence)-level backbone network, namely the Abilene network \cite{abilene}.  The Abilene network has been studied and discussed in the research literature.  Its network topology, traffic dataset, and routing information are available in the public domain \cite{dataset}.  In particular, Abilene is a public academic network in the US with 11 nodes interconnected by OC192, 9.92 Gbits/s links. (Abilene actually has another secondary core router at Atlanta, but it only connects to the primary Atlanta core router and has much less traffic.  To simplify the topology, we merged this secondary core router into the primary Atlanta core router, including all of its traffic.)

To evaluate the Abilene network, we use real traffic matrices that have been collected by a third party \cite{dataset} in our simulations.  We also use these traffic matrices in our experiments to derive the circuit configurations using our proposed network optimization algorithms.  A traffic matrix consists of the requested traffic rates for every source-destination pair within a 5-minute interval.  Therefore, these traffic matrices provide a snapshot of real total demand offerings between each IE pair in the Abilene network every five minutes.  The traffic matrices are derived based on the flow information collected from key locations of a network by traffic monitors, such as Netflow~\cite{netflow}.  Then the flow information is transformed into the demand rate of each IE pair in a traffic matrix based on the routing information in the network.  We collected the traffic matrices in each network for an extended period of time to represent the historical traffic measurements and simulation traffic load.  The detail information of the traffic matrices used is summarized in Table~\ref{table:tm}.

\begin{table}
\caption{Traffic data for Abilene.}
\begin{center}
\begin{tabular}{|c|c|c|c|c|} \hline
Network	& Collection	& Historical Traffic	& Test Traffic  	& Time		\\
	& Period	& for Allocation	& for Evaluation	& Interval	\\ \hline
Abilene	& 05/01/04	& 05/01/04		& 06/19/04		& 5 min		\\
	& to 07/02/04	& to 06/18/04		& to 07/02/04		& 		\\ \hline
\end{tabular}
\label{table:tm}
\end{center}
\end{table}

In particular, for \emph{history-based} circuit allocation, as described in \S\ref{sec:historical}, we use the historical traffic patterns during a particular time of day (3:00-3:30pm on a Wednesday) over a 7 weeks period from 05/01/2004 to 06/18/2004.  Since the dataset offers the traffic matrices at 5-minute intervals, each IE pair has 42 historical traffic data points across the analyzed period.  For our evaluations, we simulated the network traffic at the same time of day (3:00-3:30pm on a Wednesday) in the following two weeks from 06/19/2004 to 07/02/2004.  This gives another 12 traffic matrices for evaluation.

For \emph{real-time-based} circuit allocation, as described in \S\ref{sec:realtime}, we also use the 12 traffic matrices during the two weeks from 06/19/2004 to 07/02/2004 for evaluation.  For each of the 12 test traffic matrices, we interpolate the test traffic matrix with the test traffic matrix from the previous 5-minute interval, and we use this interpolated traffic matrix to define the current measured traffic rate $r_{k\ell}$ in the utility function $\phi_{k\ell}(T_{k\ell})$ shown in~(\ref{e-realtime-util}).

\subsection{Modeling Traffic Statistics}
\label{sec:model}

\begin{figure}
\centering
\includegraphics[width=\figurewidth]{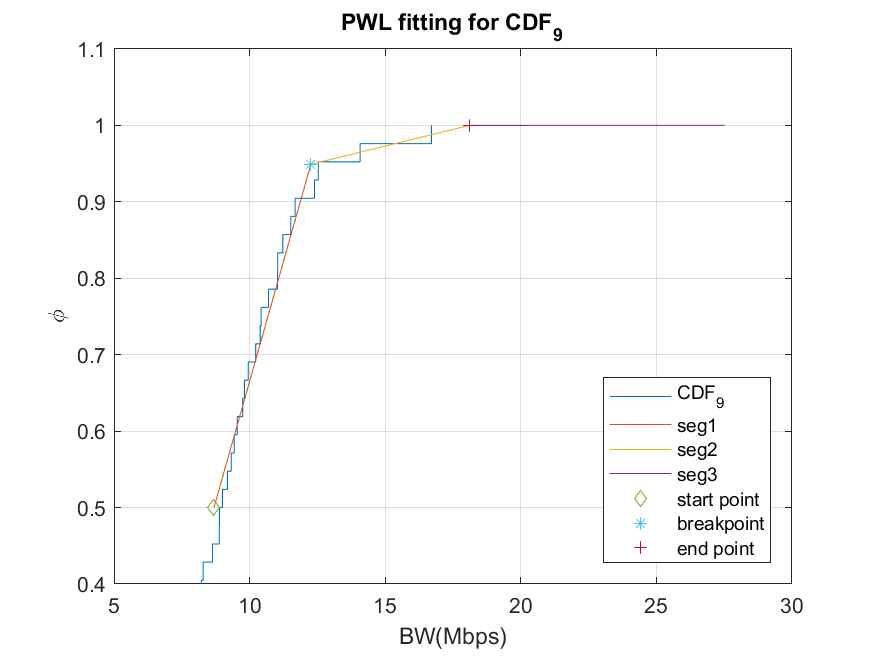}
\caption{An example PWL curve fitting of the historical traffic CDF for the flow from Atlanta to Seattle. }
\label{fig:fit_example}
\end{figure}

As discussed in \S\ref{sec:historical}, for each IE pair, we model the distribution of historical traffic patterns by an empirical cumulative distribution function (CDF).  In particular, for each IE pair, we use the collection of historical traffic data points  $\mbox{\bf r}_{k\ell} = \{r_{k\ell}(1), r_{k\ell}(2), \ldots, r_{k\ell}(t)\}$ to define a corresponding empirical CDF $\Phi_{k\ell}: \reals_+ \to [0,1]$, as shown in~(\ref{e-cdf}), and we use curve fitting to fit the empirical CDF data points $\Phi_{k\ell}(r_{k\ell}(i))$ to derive an increasing concave approximation function $\phi_{k\ell}(T_{k\ell})$.  As noted in \S\ref{sec:historical}, the CDF of a historical traffic distribution should be concave above the median traffic level.  This is because the probability density function (PDF) of traffic should be decreasing above the median level.  Therefore, we can accurately approximate the empirical CDF as a concave function by
curve fitting to those empirical CDF data points at or above the median data point for all IE pairs.

In our evaluations, we use a piecewise linear (PWL) curve fitting to approximate the empirical CDF.  Fig.~\ref{fig:fit_example} shows an example a PWL curve fitting for the IE pair traffic from Atlanta to Seattle.  In particular, the PWL curve shown corresponds to the empirical CDF of the 42 data points collected over the 7 weeks period between 05/01/2004 and 06/18/2004.  The PWL curve is fitted to all data points
at or above the median level.
In our experiments, we used CVXPY to implement a piecewise-linear curve fitting approach based on least-square fitting to a fixed number of segments (\eg, 3 segments).  More sophisticated piecewise linear curve fitting approaches (\eg, \cite{magnani2009convex}) can be used as well.

\subsection{Circuit Allocation}
\label{alloc}

We performed the circuit allocation for all 11 cities in the Abilene network, corresponding to 110 IE pairs, by solving the convex optimization problem ({e-network-opt}) in \S\ref{sec:allocate}.   For real-time-based circuit allocation, we use $\phi(T)$ as defined in \S\ref{sec:realtime}, and for history-based circuit allocation, we use the PWL curve fitting approach shown above to derive $\phi(T)$, as discussed in \S\ref{sec:historical}.  For $\alpha$-fairness~(\cf~(\ref{e-alpha})), we assume $\alpha=2$.
To solve the convex optimization, we use CVXPY \cite{cvxpy} with MOSEK \cite{mosek}.

\subsection{Re-Routing over Circuits for Adaptation}
\label{sec:reroute}

Although both our real-time-based and history-based circuit allocation algorithms aim to allocate circuit capacities so that actual traffic can be carried directly by the allocated circuits, traffic fluctuations or unexpected traffic changes can lead to inadequate capacities along direct circuits. One way to handle the excess traffic is to adaptively re-route the excess traffic over circuits with spare capacity. Since our circuit allocation algorithms are designed to create direct circuits between every IE-pairs, the logical network topology becomes a fully-connected mesh.

\begin{figure}
    \centering
    \includegraphics[width=\figurewidth]{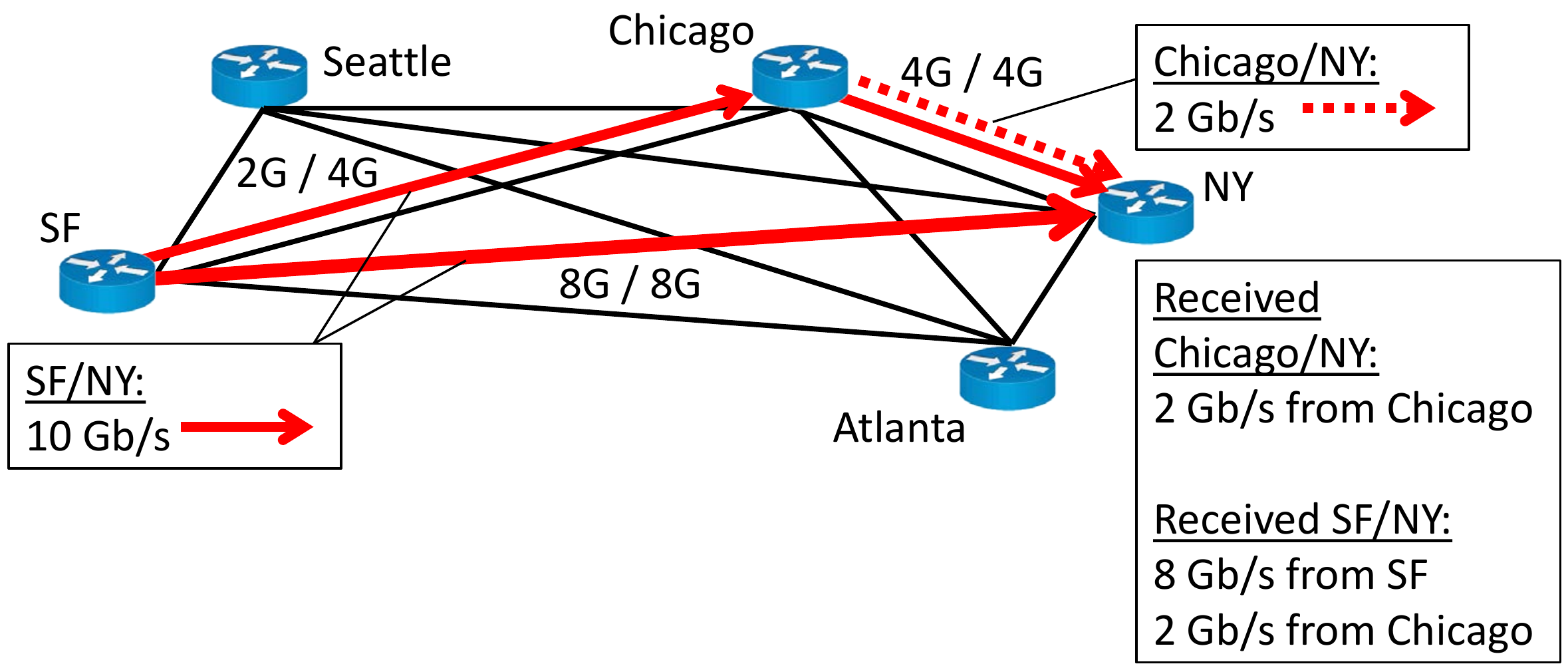}
    \caption{When excess traffic occurs from SF to NY, we can re-route it
             using the residual circuit capacity of the path through for
             example Chicago.}
    \label{fig:reroute}
\end{figure}

Consider the example depicted in Fig.~\ref{fig:reroute}. Suppose the circuit capacity from SF (San Francisco) to NY (New York) is 8 Gb/s, and suppose the circuit capacities from SF to Chicago and Chicago to NY are both 4 Gb/s.  Normally, we expect a circuit to have enough capacity for its direct traffic. For example, in Fig.~\ref{fig:reroute}, given 2 Gb/s of traffic from Chicago to NY, all of its traffic can be directly sent through the network using the circuit from Chicago to NY.  However, suppose we have a 10 Gb/s burst of traffic from SF to NY, then there would be 2 Gb/s of excess traffic because the circuit capacity from SF to NY is only 8 Gb/s. When this occurs, an adaptive re-routing mechanism can be triggered to re-route the 2 Gb/s of excess traffic over alternative circuit routes, for example through Chicago by the utilizing the residual circuit capacity available along SF-Chicago and Chicago-NY.

As can be seen from this example, with the help of adaptive re-routing, we can increase network throughput without the need to create new circuits on-the-fly. Although this adaptive re-routing does rely on electronic routing at intermediate nodes, it is only used as a secondary mechanism to handle excess traffic. The majority of traffic is still expected to be carried by the corresponding direct circuits. Therefore, the route processing portion of a unified circuit/packet switch can remain simple.

In our experiments, we consider two re-routing methods.  One is based on a variant of the well-known backpressure-based re-routing algorithm \cite{Yin2017} that guarantees optimal re-routing.  In the modified version of the backpressure-based re-routing algorithm, a unified switch maintains a separate queue for packets for each destination, and it transmits packets on the direct circuit as long as the circuit has sufficient capacity.  Insufficient capacity is detected when the queue of packets for the direct circuit exceeds some threshold $L_{max}$.  When this occurs, packets are re-routed using the backpressure algorithm.  The re-routing is optimal in the sense that if a traffic pattern can be handled by re-routing over the logical fully-meshed network of circuits, then the re-routing algorithm is guaranteed to succeed in re-routing all traffic to their destinations. In \S\ref{eval_routing}, we refer to this re-routing approach as ``OptRR'' for optimal re-routing.

Alternatively, we also consider a simple greedy re-routing algorithm that simply re-routes the excess traffic over the outgoing circuit with the most residual capacity.  Suppose $T_{m\ell}$ is the circuit capacity allocated to the circuit from the current node $\ell$ to node $m$, and suppose $\mu_{m\ell}$ is the measured rate of traffic sent on the circuit from the current node $\ell$ to node $m$ in the current measurement interval.  Then the amount of ``residual capacity'' on the circuit from the current node $\ell$ to $m$ is simply $T_{m\ell} - \mu_{m\ell}$.  A simple greedy algorithm is just to re-route traffic to node $m$ via the circuit to node $m$ with the most residual capacity rather than directly to destination $k$.  This greedy approach only requires information that can be measured locally, but it is not optimal.  We include this re-routing method in our experiments to show that even this simple approach is effective with our circuit allocation methods.  In \S\ref{eval_routing}, we refer to this re-routing approach as ``GreedyRR'' for greedy re-routing.

\section{Experimental Results}
\label{sec:results}

In this section, we first evaluate the performance of our history-based circuit allocation algorithm in terms of what fraction of the historical traffic patterns that the allocated circuits can handle as well as the fraction of test traffic patterns that the allocated circuits can handle.  We then compare the performance of circuit-switching approaches using our circuit allocation methods, namely the real-time-based and history-based circuit allocation approaches, with a conventional packet-routing algorithm, OSPF \cite{ospf} in \S\ref{eval_routing}.  We extend our circuit-switching approaches with adaptive re-routing in cases when the circuit capacity is not enough, as discussed in \S\ref{sec:reroute}.  This re-routing approach is also evaluated in \S\ref{eval_routing}.  Our evaluations show that our circuit allocation algorithms can indeed accommodate most of the actual traffic, and adaptive re-routing over the allocated circuits can effectively accommodate excess traffic even under heavy traffic loads.

\subsection{Evaluation of History-Based Circuit Allocation}
\label{eval_circuit_allocation}

\begin{figure}
  \centering
  \subfigure[Historical traffic.]{
    \label{fig:achievableHis}
    \includegraphics[width=\figurewidth]{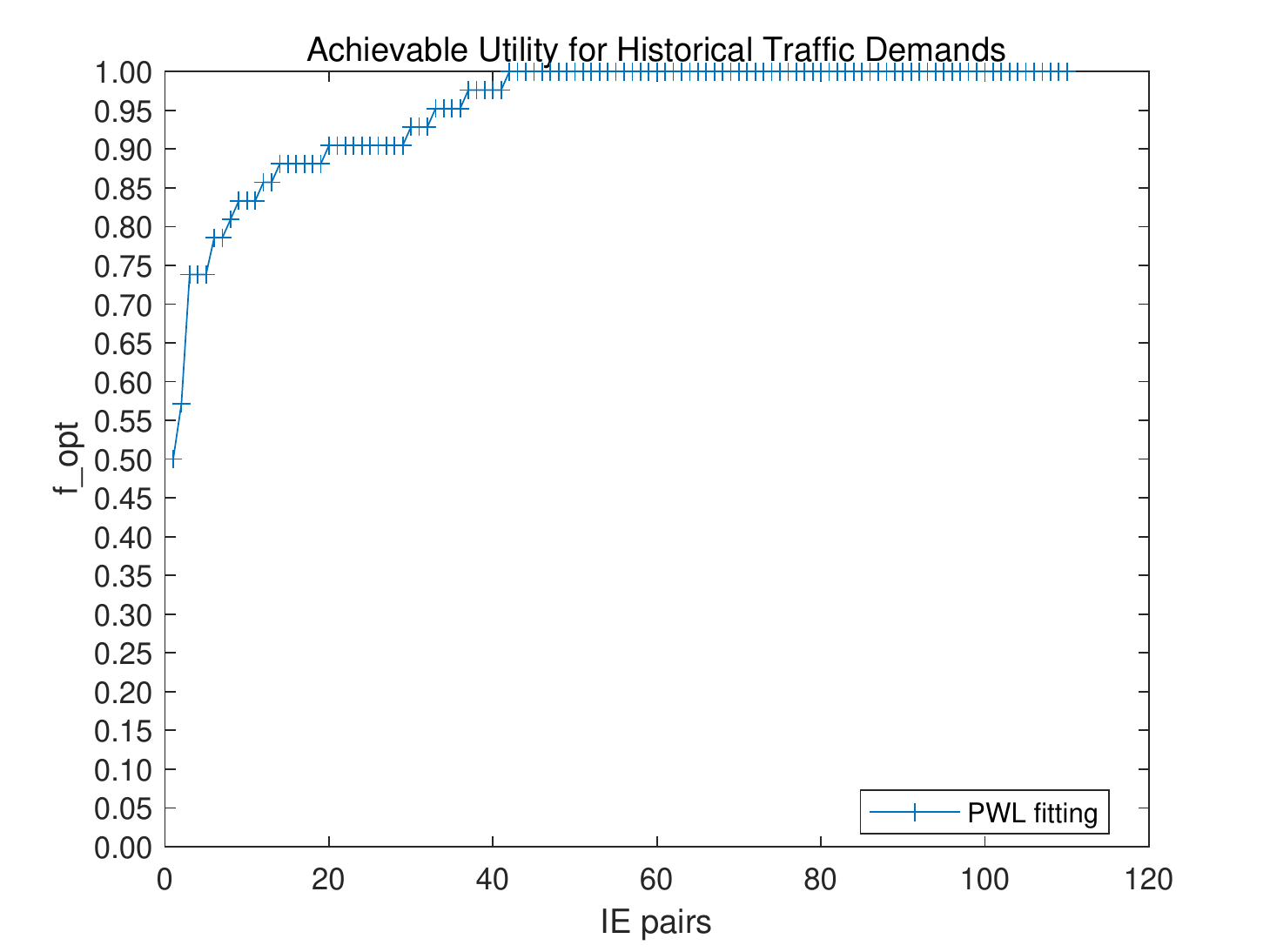}
  }
  \subfigure[Test traffic.]{
    \label{fig:achievableTest}
    \includegraphics[width=\figurewidth]{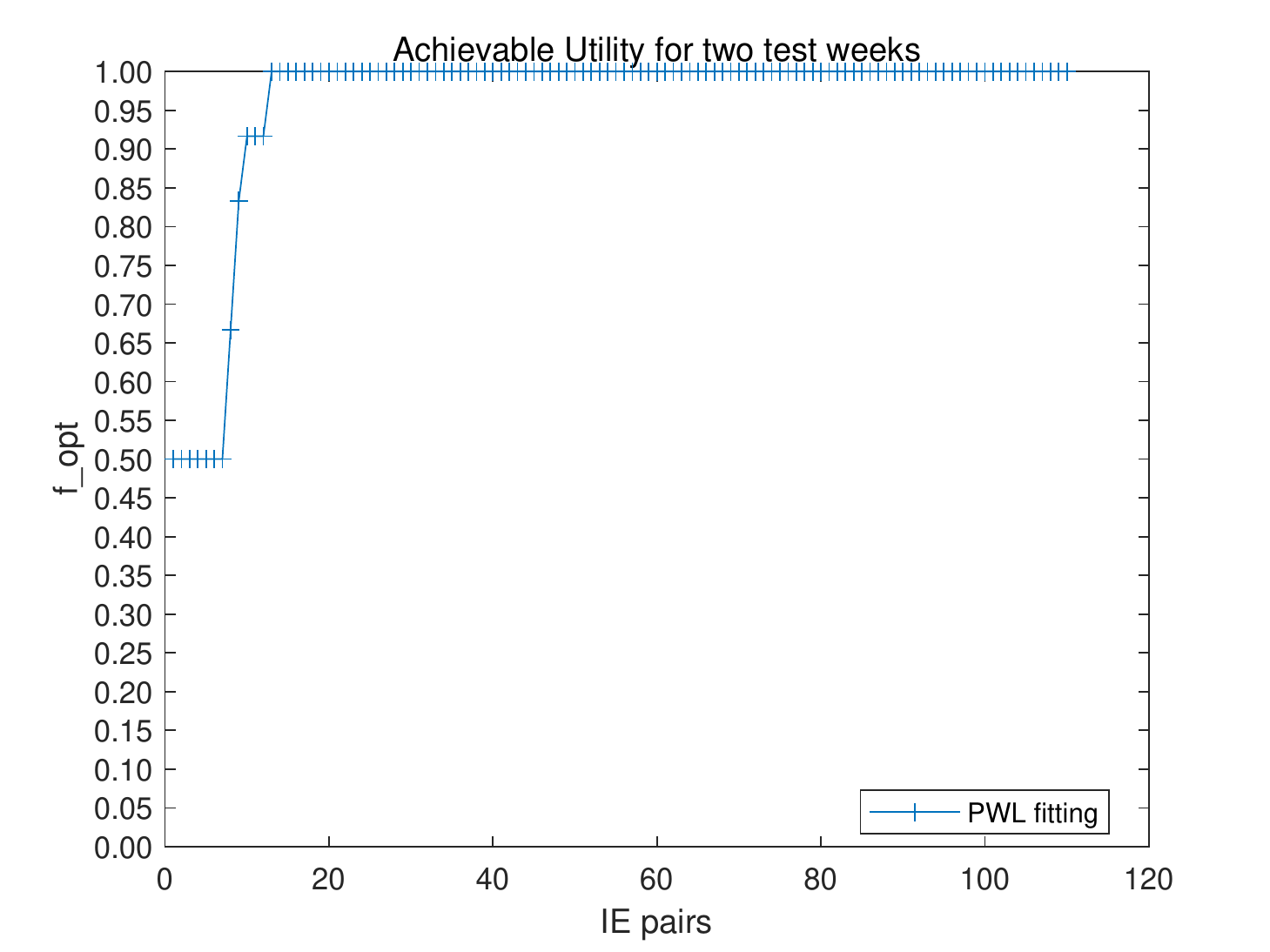}
   }
   \caption{Achievable utility for historical and test traffic demands}
   \label{fig:achievable}
\end{figure}

\begin{figure}
  \centering
  \includegraphics[width=\figurewidth]{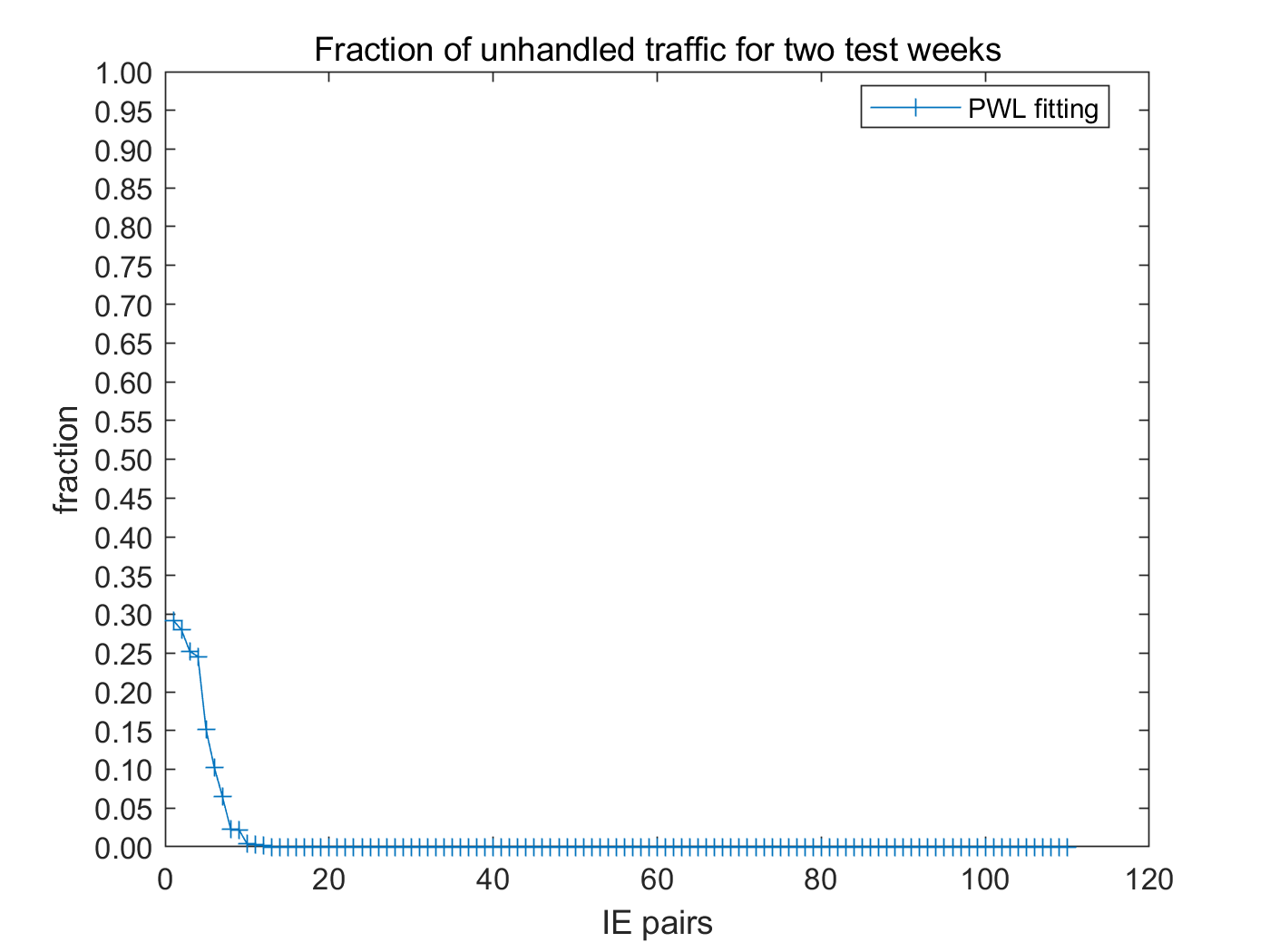}
  \caption{Fraction of unhandled traffic for two test weeks}
  \label{fig:unhandled}
\end{figure}

%

Figure~\ref{fig:achievable} shows the fractions of the data points whose traffic demands may be accommodated by the optical circuit allocation solved by the convex optimization problem solver for all $110$ IE pairs.  The achievable $f_{opt}$ on the Y-axis means the allocation $T_{k\ell}$ is no less than the fraction $f_{opt}$ of the traffic data points for IE pair $(k, \ell)$.  For example, $f_{opt} = 0.9$ for an IE pair means
$T_{k\ell}$ is greater or equal to 90\% of the traffic data points for that IE pair $(k, \ell)$ at a given time, and $f_{opt} = 1$ means $T_{k\ell}$ can accommodate all of the traffic data points for that IE pair.
In particular, figure~\ref{fig:achievableHis} shows the coverage of the historical traffic patterns, and figure~\ref{fig:achievableTest} shows the coverage of the test traffic patterns.

As can be seen from figure~\ref{fig:achievableHis}, the circuit configuration is able to accommodate all historical traffic data points for more than two thirds of all 110 IE pairs.  The smallest fraction occurs at $0.5$, and that is for only one flow.  When the circuit configuration is applied to the two test weeks, figure~\ref{fig:achievableTest} shows that this configuration can accommodate all data points for more than 80\% IE pairs.

Figure~\ref{fig:achievable} only considers the traffic demands that are strictly lower than the optical circuit bandwidth. If a traffic demand is slightly higher than the given circuit bandwidth, the circuit allocation is considered to fail to accommodate that data point.  As can be seen from figure~\ref{fig:achievableTest}, the circuit allocation of some IE pairs fails to accommodate half of the traffic demand data points.  However, the actual unhandled traffic in this case may be small.  Therefore, figure~\ref{fig:unhandled} is used to show the amount of unhandled traffic for the test weeks when our circuit allocation is used.

As can be seen from figure~\ref{fig:unhandled}, our history-based circuit allocation can accommodate all traffic demands for 90\% IE pairs in the test weeks, and only less than 30\% traffic from the worst-case IE pair is unhandled by the allocated bandwidth.


\subsection{Performance Evaluations}
\label{eval_routing}

To evaluate the performance of our circuit allocation approach, we compare the following:
\begin{itemize}
\item \emph{OSPF}: This is conventional packet routing over the Abilene network in which the routing paths are determined using the Open Shortest Path First (OSPF) protocol \cite{ospf}, which is used for packet routing.  The routing paths are based on Dijkstra's single shortest path algorithm.  This conventional approach serves as a baseline for our evaluations.

\item \emph{RT}: The plots labeled ``RT'' correspond to our real-time-based circuit allocation algorithm for optical circuit-switching.  In particular, we consider three cases.  The first case corresponds to circuit-switching without re-routing.  Here, traffic is also simply sent directly over a fully-meshed network in one logical hop, whose circuit capacities are determined by the algorithm described in \S\ref{sec:realtime}.  This is labeled as ``RT-NoRR.''  The other two cases correspond to the two methods of re-routing, as discussed in \S\ref{sec:reroute}.  ``RT-GreedyRR'' corresponds to greedy re-routing based on residual capacity, and ``RT-OptRR'' corresponds to optimal re-routing based on a modified version of the backpressure algorithm for re-routing \cite{Yin2017}.

\item \emph{HIST}: The plots labeled ``HIST'' correspond to our history-based circuit allocation for optical circuit-switching.  ``HIST-NoRR,'' ``HIST-GreedyRR,'' and ``HIST-OptRR'' correspond to the cases of no re-routing, greedy re-routing based on residual capacity, and optimal re-routing based on the backpressure algorithm.
\end{itemize}
For each method, we simulated the network traffic during the two weeks from 06/19/2004 to 07/02/2004 at the same time of day (3:00-3:30pm on a Wednesday).  This provides 12 traffic matrices for evaluation. We measured the drop rates, router hops, router loads, and the percentage of packets being routed.  The results presented are averaged over the 12 test cases.  To demonstrate the performance of our algorithms under high traffic loads, we normalized the traffic by scaling up the traffic loads until OSPF routing begins to drop packets. That is, a normalized traffic load of 1.0 is the intensity of traffic that causes the network to saturate when conventional packet switching with OSPF is used. To test the robustness of our circuit allocation approaches, we further amplify the traffic intensity beyond this saturation point to see how much more traffic our circuit allocation approaches with re-routing can handle.

\begin{figure}
  \centering
  \includegraphics[width=\figurewidth]{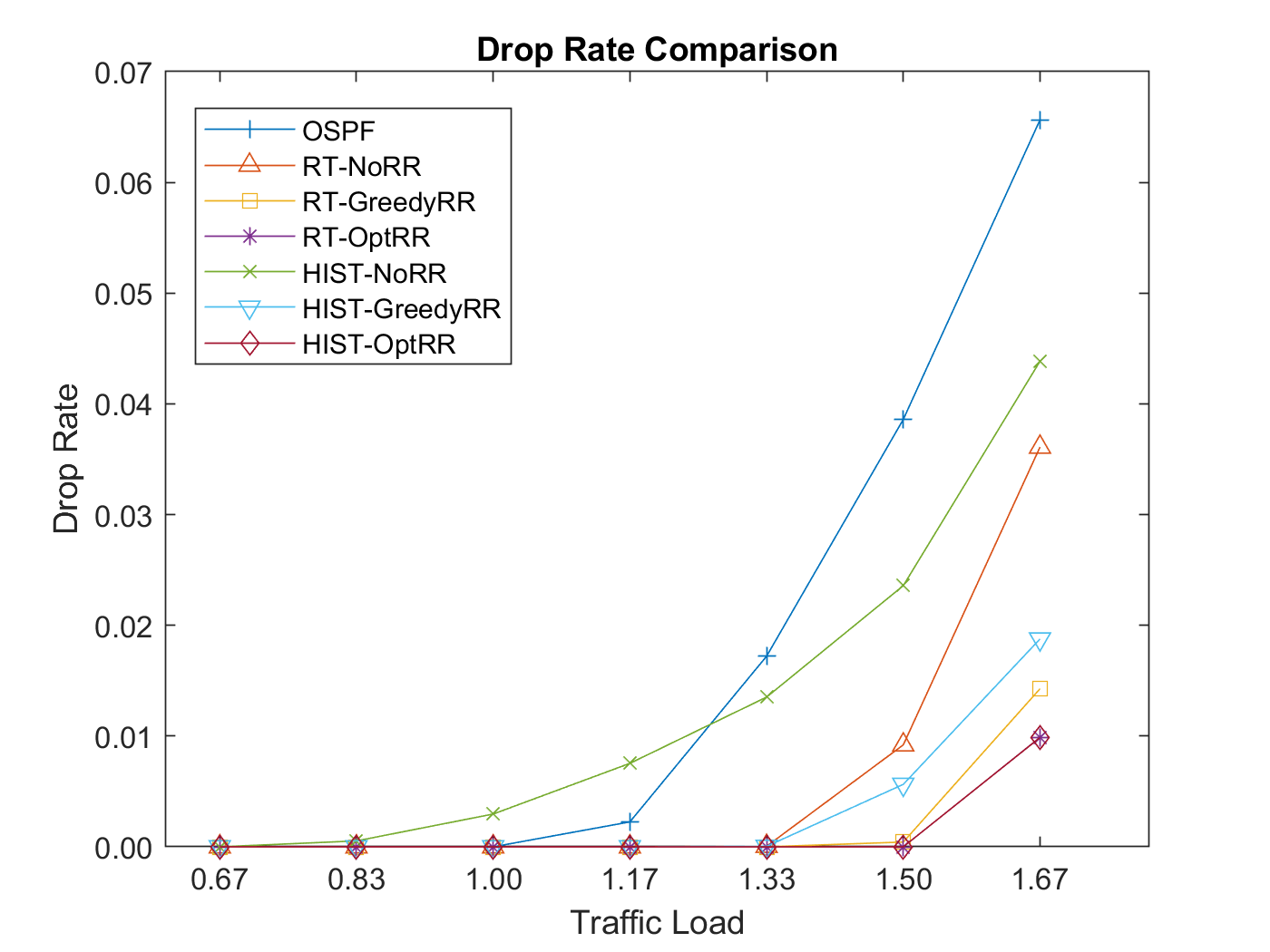}
  \caption{Drop rate comparison.}
  \label{fig:drop_rate}
\end{figure}

\begin{figure}
  \centering
  \includegraphics[width=\figurewidth]{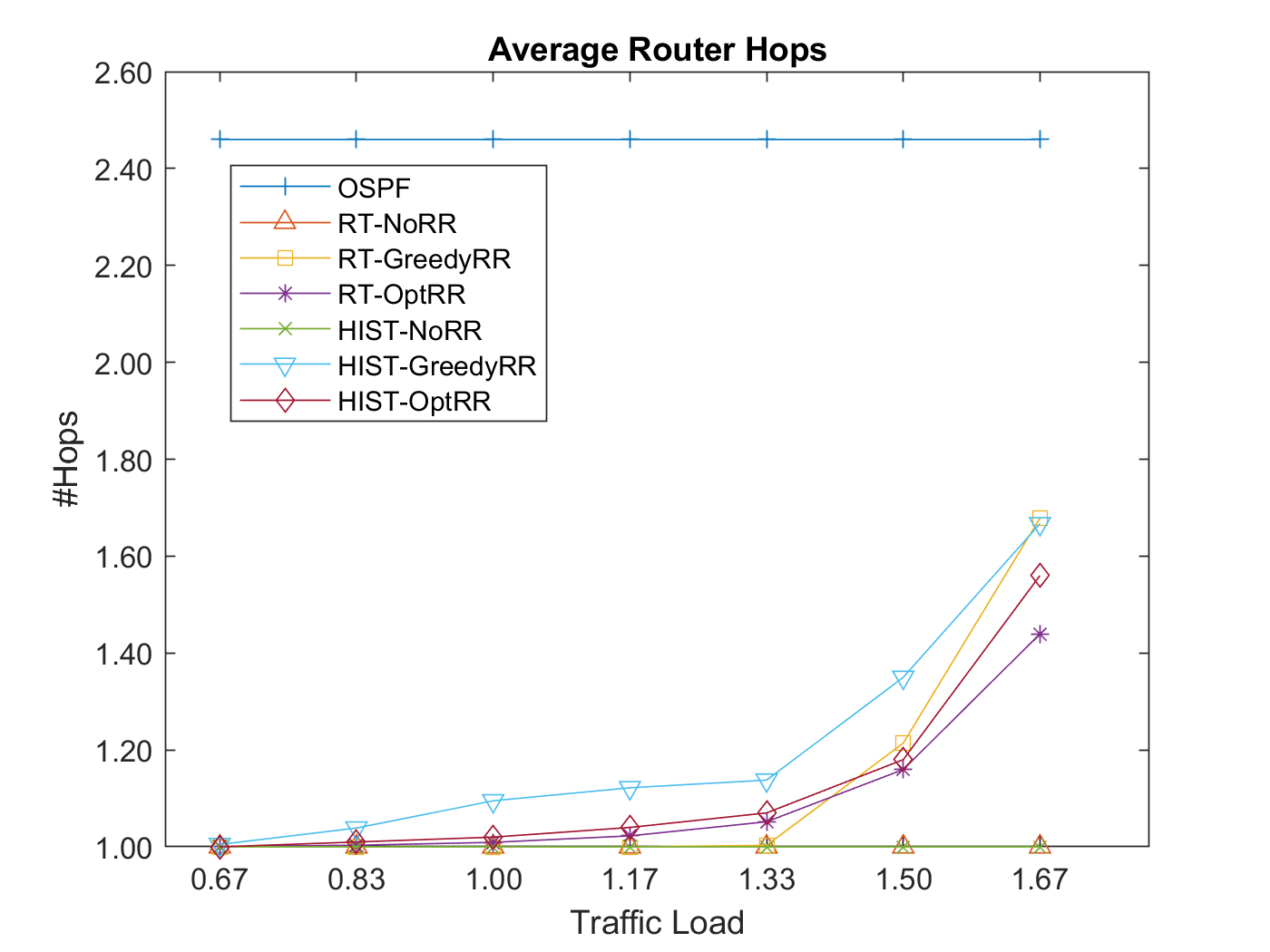}
  \caption{Average router hops.}
  \label{fig:router_hops}
\end{figure}

\begin{figure}
  \centering
  \includegraphics[width=\figurewidth]{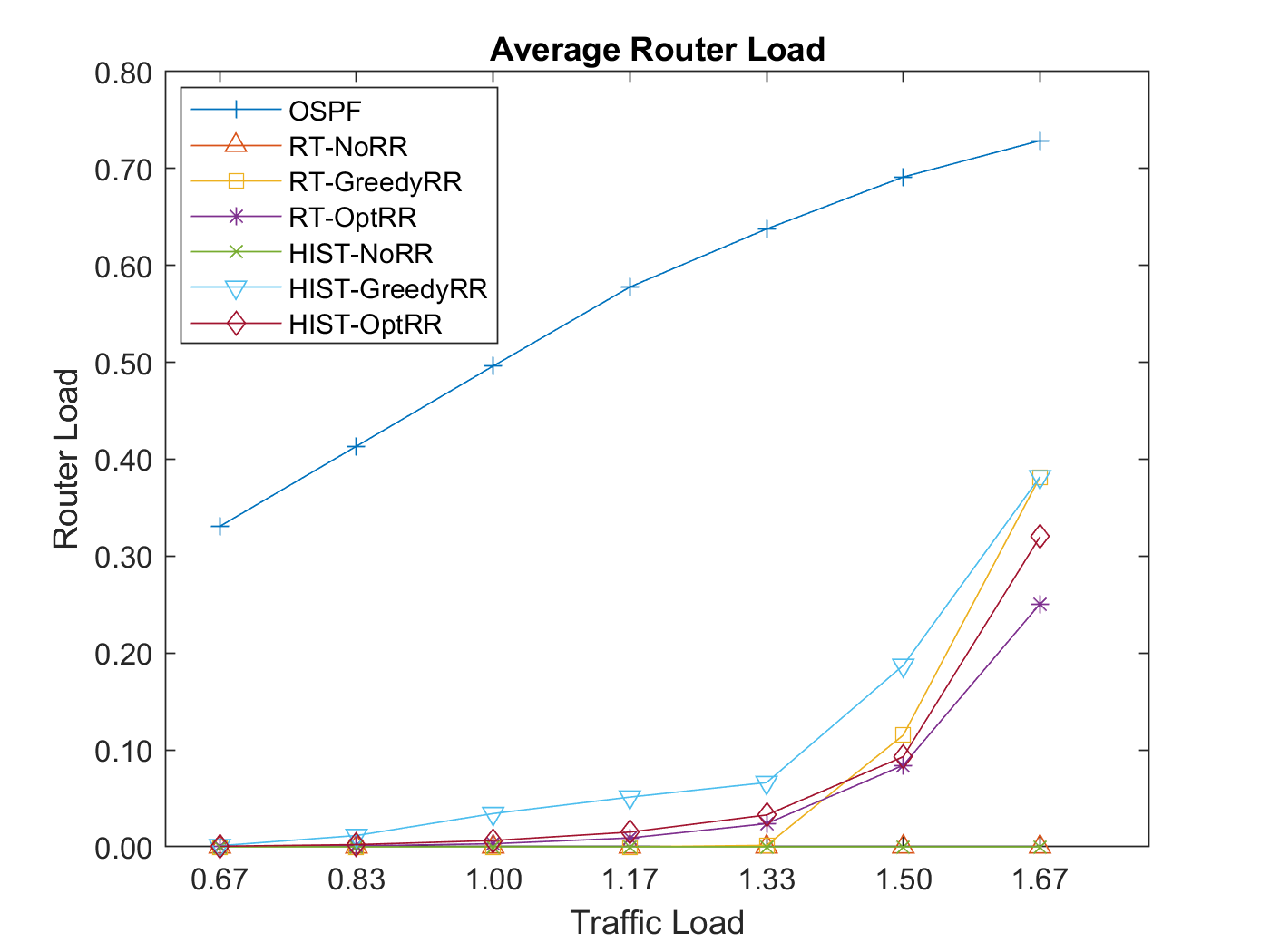}
  \caption{Average router load.}
  \label{fig:router_load}
\end{figure}

\begin{figure}
  \centering
  \includegraphics[width=\figurewidth]{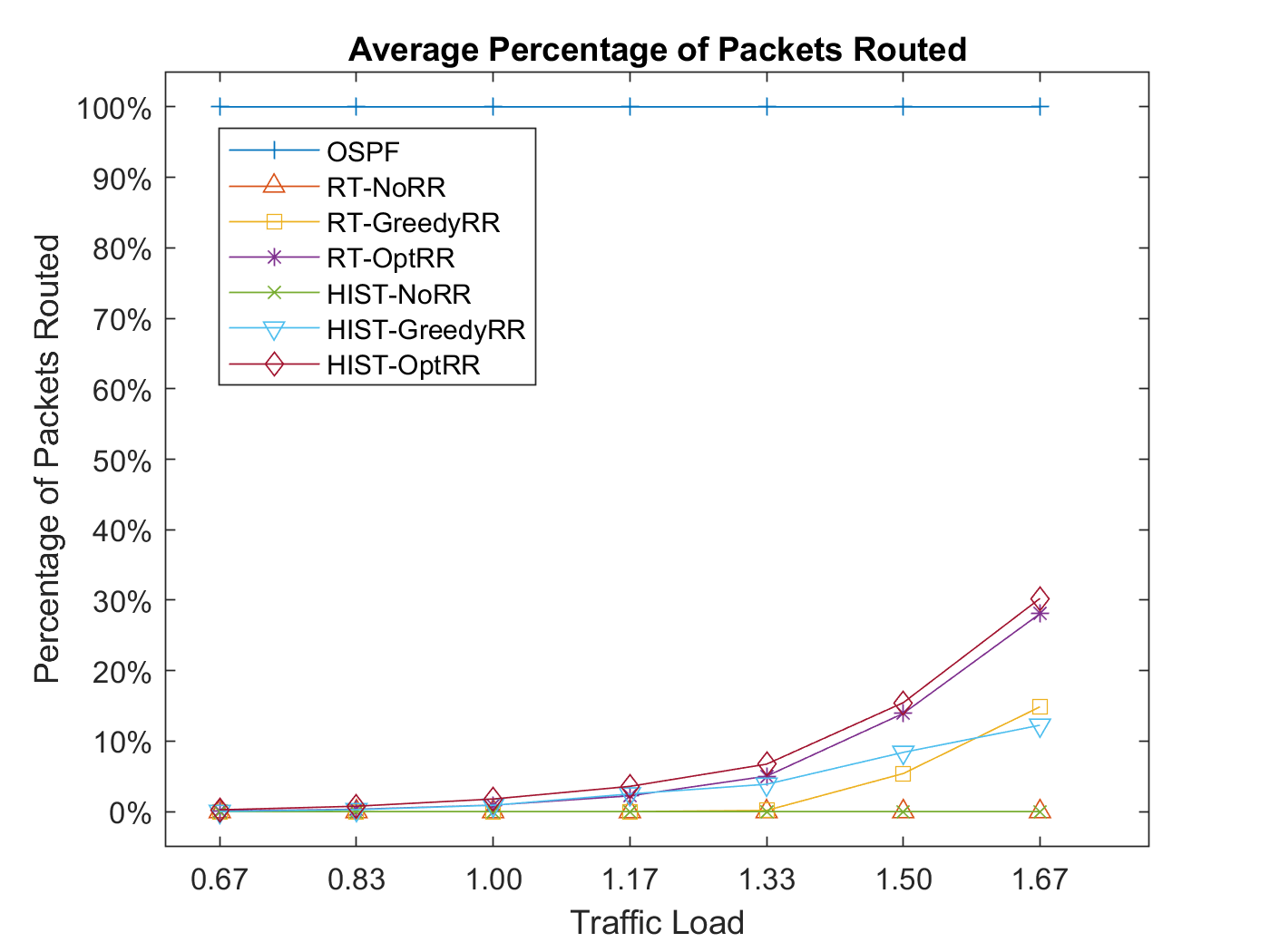}
  \caption{Average percentage of packets routed.}
  \label{fig:packets_routed}
\end{figure}

Figure~\ref{fig:drop_rate} compares drop rates among real-time-based allocation (RT-NoRR, RT-GreedyRR, and RT-OptRR), history-based allocation (HIST-NoRR, HIST-GreedyRR, and HIST-OptRR), and OSPF. The suffixes NoRR, GreedyRR, and OptRR correspond to no re-routing, greedy re-routing, and optimal re-routing, respectively, as discussed in \S\ref{sec:reroute}. The X-axis represents traffic loads which are normalized to the load where OSPF begins to drop packets. As we can see, compared with OSPF, our ``re-routing'' approaches (RT-GreedyRR, RT-OptRR and HIST-OptRR) can handle 50\% more traffic without dropping packets. Even with greedy re-routing, our history-based approach (HIST-GreedyRR) can handle 33\% more traffic without dropping packets. With ``no re-routing,'' our history-based circuit allocation approach (HIST-NoRR) has a negligible drop rate (0.00296) at the normalized load of 1.0 while OSPF has none, but HIST-NoRR has a lower drop rate as traffic continues to scale. As expected, with ``no re-routing,'' our real-time-based approach (RT-NoRR) achieves significantly better results than the history-based approach (HIST-NoRR) because the circuit allocations are based on real-time traffic measurements. Even without re-routing, our real-time-based approach (RT-NoRR) can handle 33\% more traffic than OSPF without dropping packets. The reason for the higher throughput is because OSPF always route along the shortest path, whereas our circuits can be configured across multiple paths, optimized to real-time traffic measurements. Although the real-time-based circuit allocation approach performs better, both the real-time-based and the history-based approaches are important, depending on whether or not the deployment scenario allows for real-time measurements and frequent updates or not.  The Internet, as implemented today, does not have real-time measurements or allow for frequent dynamic updates, but emerging software defined networking scenarios would provide for that.  Our optimization framework supports both settings.

Figure~\ref{fig:router_hops} shows the number of router hops a packet needs to go through until it reaches its destination.  With no re-routing, the real-time-based (RT-NoRR) and history-based (HIST-NoRR) approaches are both always 1 hop over the direct optical circuit, whereas OSPF routing averages 2.46 hops independent of load. With optimal re-routing, both real-time-based (RT-OptRR) and history-based (HIST-OptRR) approaches require very little re-routing for loads up to 1.5X. With greedy re-routing, the real-time-based approach (RT-GreedyRR) also requires very little re-routing for loads up to 1.5X, whereas the history-based approach (HIST-GreedyRR) requires very little re-routing for loads up to 1.33X. As expected, these results show that the real-time-based approach is more accurate than the history-based approach, but better re-routing can compensate for the difference. At higher normalized loads, less re-routing is required when real-time-based allocation is used together with optimal re-routing.


Figure~\ref{fig:router_load} shows the average router load among real-time-based allocation (RT-NoRR, RT-GreedyRR, and RT-OptRR), history-based allocation (HIST-NoRR, HIST-GreedyRR, and HIST-OptRR), and OSPF. With ``no re-routing'' (RT-NoRR and HIST-NoRR), all packets go over direct optical circuits and therefore these approaches have 0 electronic router load. For OSPF, all packets are handled by electronic routers, so as expected, the router load increases with traffic load. When optimal re-routing is employed with both the real-time-based (RT-OptRR) and history-based (HIST-OptRR) approaches, we see that most packets go over direct optical circuits until 1.33X traffic load; after that, the electronic router load increases as more packets get re-routed. As with the average number of hops, we see that the history-based approach (HIST-OptRR) requires a higher router load than the real-time-based approach (RT-OptRR) when the normalized traffic load is increased to 1.67X. This is mostly due to the fact that when packets are re-routed, they go through a high number of intermediate nodes (a higher number of hops). When greedy re-routing is employed, router loads are comparable at 1.67X normalized loads for both real-time-based (RT-GreedyRR) and history-based (HIST-GreedyRR) allocation approaches.

Finally, figure~\ref{fig:packets_routed} shows the average percentage of packets that require routing.
For OSPF, 100\% of the packets are routed. With ``no re-routing'' (RT-NoRR and HIST-NoRR), none of the packets are routed since they all go over direct optical circuits. With ``re-routing'' (RT-GreedyRR, RT-OptRR, HIST-GreedyRR, and HIST-OptRR), most packets go over direct optical circuits until 1.33X traffic load (less than 10\% of packet gets re-routed at this load); after that, an increasing percentage of packets get re-routed. When packets have to re-routed, the optimal re-routing approaches (RT-OptRR and HIST-OptRR) route a higher percentage of packets, but most of the time by a fewer a number of hops in comparison with greedy re-routing (RT-GreedyRR and HIST-GreedyRR).


\section{Additional Related Work}
\label{sec:related}

Previous approaches have been proposed for the allocation of circuits to handle specific traffic matrices 
\cite{banerjee00,ramamurthy00,tornatore02}. Our work is different in several ways. First, in the history-based allocation setting, our formulation takes into consideration the statistical daily traffic variations observed in past measurements and the probability of traffic demands given their statistical distribution of occurrence in past measurements. In our formulation, the allocated circuits do not necessarily provide sufficient circuit capacities for supporting all the traffic matrices captured in the historical data sets. Instead, our problem is formulated as a utility max-min fair circuit allocation problem that aims to maximize the acceptance probability of the expected traffic demand by using the cumulative distribution function over the historical data sets as the objective function. Our solution allocates all available network resources across multiple paths to provide as much headroom as possible. Since our solution does not rely on an online dynamic circuit creation mechanism, there is no need to leave behind network resources for establishing new circuits.

Second, even in the case of our real-time-based allocation setting, our problem formulation allows for the actual traffic in the next period to be different from the current measurement period, and we fully allocate all network resources allow for some fluctuations in traffic rates. This setting is formulated as a weighted max-min fair circuit allocation problem. Our convex optimization approach makes it possible to solve both problems in a unified framework.

There have also been prior work on weighted max-min fair allocation and utility max-min fair allocation for bandwidth allocation problems, but they either only considered the single-path case \cite{cao99,rubenstein02,radunovic07} or provided approximate solutions \cite{allalouf05,coplar,ummf} in the multipath case based on a binary search of achievable utilities. Our approach in this paper is different in that the problems are solved as a single convex optimization problem, including the modeling of historical traffic distributions as concave functions.

\section{Conclusion}
\label{sec:conc}

In this paper, we considered circuit allocation problems for unified packet and circuit switched networks. We proposed a novel convex optimization framework based on a new destination-based multicommodity flow formulation for the allocation of circuits in such unified networks. In particular, we consider two deployment settings, one based on real-time traffic monitoring, and the other relying upon history-based traffic predictions. In both cases, we formulate global network optimization objectives as concave functions that capture the fair sharing of network capacity among competing traffic flows. The convexity of our problem formulations ensures globally optimal solutions.

\bibliographystyle{plain}
\bibliography{ref}

\end{document}